\documentstyle[12pt,fullpage,subeqnarray]{article}
\def\be{\begin{equation}}
\def\ee{\end{equation}}
\def\eq#1{(\ref{#1})}
\begin{document}
\begin{flushright}
TIFR/TH/95-09
\end{flushright}
\bigskip
\begin{center}
{\large \bf A VIEW OF 2-DIM. STRING THEORY AND BLACK
HOLES\footnote{Based on the talk given at the `International
Colloquium on Modern Quantum Field Theory', Bombay, January 1994}} \\
\bigskip
{\large Spenta R. Wadia \\
Tata Institute of Fundamental Research, \\ Homi Bhabha Road,
Bombay 400 005, INDIA} \\
{\em e-mail: wadia@theory.tifr.res.in}
\end{center}
\abstract{We present a brief overview of 2-dim. string theory and its
connection to the theory of non-relativistic fermions in one
dimension.  We emphasize (i) the role of $W_\infty$ algebra and (ii)
the modelling of some aspects of 2-dim. black hole physics using the phase
space representation of the fermi fluid.}

\vskip 1cm
\setlength{\baselineskip}{14pt}

\noindent {\bf 0.~ Introduction}
\vskip 0.5cm

There are many issues of physical interest in string theory that
require a non-perturbative formulation; for example black holes,
supersymmetry breaking, quark lepton masses are a few of the
phenomena of interest.  Presently such a formulation is not
available to us in 4 dimensions.  Not only do we not know the laws of
string theory in 4-dim. but we do not even know its true microscopic
degrees of freedom.  It is in these circumstances that toy models
become important because their formulation and solution may shed some
light on the more realistic issues of string theory which we would
like to address.

An important model that has been much studied is 2-dim. string theory.
There are several reasons for this:

\begin{enumerate}
\item[{(i)}] Firstly 2-dim. string theory is non-trivial in content.
Its weak coupling (low energy) spectrum has a massless particle.
There is a non-trivial $S$-matrix of the massless particles.  Besides
that it also has a discrete infinity of backgrounds which are the
remnants of the massive string modes in higher dimensions.

\item[{(ii)}] The low energy limit (in the $\sigma$-model approach)
has a black hole solution which is characterized by a mass.

\item[{(iii)}] There is a matrix model formulation of the theory that
in principle defines the theory non-perturbatively for all values of
the coupling constant.  The matrix model is exactly formulated as a
system of non-relativistic fermions in 1-dimension (the full real
line) in an inverted harmonic oscillator potential.

\item[{(iv)}] The matrix model formulation has several important
implications:
\begin{enumerate}
\item[{a)}] Firstly it enables computations which can be performed to
all orders in perturbation theory.

\item[{b)}] It indicates in a simple way the existence of
non-perturbative effects that go as $e^{- {1 \over g_{str}}}$ which are
characteristic of string theory.

\item[{c)}] There are an infinite number of conservation laws.  The
phase space of the theory is characterized as a non-linear
representation of $W_\infty$ algebra.

\item[{d)}] There is a real possibility of studying non-perturbative
aspects of black holes.

\end{enumerate}
\end{enumerate}

\vskip 1cm

\noindent {\bf 1.~ The Matrix Model Formulation of 2-dim. String
Theory}
\vskip 0.5cm

In the following we briefly spell out the matrix model formulation and
indicate the connections with the target space.

As is well known a discrete formulation of 2-dim. gravity leads to the
matrix model\cite{one}.  The 1-dim. matrix model describes the
coupling of 1-dim. matter to 2-dim. gravity.  The Liouville mode of
2-dim.  gravity can be identified with a space co-ordinate\cite{two}
and hence this model is in fact a string theory in a 2-dim. target
space-time.  It is in this way that the matrix model describes
2-dim. string theory.  The matrix model in turn maps into the problem
of non-relativistic fermions in 1-dim.\cite{three} In the double
scaling limit the non-interacting fermions move in an inverted
harmonic oscillator potential.\cite{four} We can write the action in
terms of the non-relativistic fermion field $\psi(x,t)$:
\begin{eqnarray}
\label{1}
S & = & \int^{+\infty}_{-\infty} dt \int^{+\infty}_{-\infty} dx
\psi^+(x,t)\: \left(i\partial_t - h_x\right)\: \psi(x,t) \nonumber \\
h_x & = & {1\over 2}\: \left(-\partial^2_x + V(x)\right) \nonumber \\
V(x) & = & -x^2 + {g_3  x^3 \over \sqrt N} + \ldots \nonumber  \\
&& \hbox{~~~} \int^{+\infty}_{-\infty} \psi^+\psi dx  =  N.
\end{eqnarray}
The double scaling limit corresponds to $N\rightarrow \infty$ and the
bare Fermi level (measured from the maximum of the potential)
$\epsilon_F \rightarrow 0$, while keeping $\mu = N\epsilon_F$ fixed.

It is natural to identify the fermions as the microscopic degrees of
freedom of the string theory. Since fermion number is held fixed, the
physical variables are those which are invariant under the $U(1)$
transformation $\psi(x,t)
\rightarrow e^{i\Theta} \psi(x,t)$.  A general set of variables with
this property are the bilocal variables\cite{five,six}:
\be
\label{2}
\phi(x,y,t) = \psi(x,t)\; \psi^+ (y,t)
\ee
They satisfy the $W_\infty$ algebra\cite{five,six,seven,eight,nine}
\begin{eqnarray}
\label{3}
\left[\phi(x,y,t), \phi(x',y',t)\right] & = & \phi(x'y,t)\; \delta
(x-y') \nonumber \\
&& -\phi(x,y',t)\;\delta(x'-y)
\end{eqnarray}
These are the Poisson brackets of our phase space. However the phase
space is not linear and there are non-linear constraints reflecting the
underlying fermion degree of freedom.  Defining $\Phi(t)$ such that
\be
\label{4}
\langle x |\Phi(t)|y \rangle = \phi (x,y,t)
\ee
the non-linear constraints are easily deduced using the fermion
anti-commutation relations,
\begin{eqnarray}
\label{5}
\Phi^2 & = & \Phi \nonumber \\
tr(1-\Phi) &=& N.
\end{eqnarray}
The equation of motion is
\be
\label{6}
i\partial_t \Phi + [\hat h,\Phi] = 0.
\ee
where $\hat h$ is the single particle operator $\hat h = {1\over
2}\:\hat p^2 + V(\hat x)$.

\vskip 1cm

\noindent {\bf 2.~ The Classical Fermi fluid}

\vskip 0.5cm

One can present an action principle and a path integral formulation
for the system of equations \eq{3},\eq{4},\eq{5}.  For details this
we refer the reader to the published literature.\cite{six,ten}
Presently we express the above formulae in terms of the operator that
describes the phase space distribution function of the fermions.  It
is basically a double fourier transform of the bilocal variable
$\phi(x,y,t)$:
\be
\label{7}
\hat u (p,q,t) = \int dx \psi^+(q-{x \over 2}, t)\: e^{-ipx}\; \psi\left(q
+ {x\over 2},\; t\right)
\ee
Let us denote the classical phase-space distribution by $u(p,q,t)$.
It is the expectation value of \eq{7} in a $W_\infty$ coherent state.
It satisfies the Liouville equation (follows from \eq{6})
\be
\label{8}
\partial_t u + p\partial_qu + q\partial_pu = 0
\ee
The constraints on $u$ are (follows from \eq{5}):
\begin{subeqnarray}
\left[\cos {1\over 2} \left(\partial_p\partial_{q'} -
\partial_{p'}\partial_q\right)\; u(p,q,t)\: u(p',q',t)\right]_{p' = p
\atop q' = q} & = & u(p,q,t) \slabel{9a} \\
\int {dpdq \over 2\pi}\: u(p,q,t) & = & N. \slabel{9b}
\end{subeqnarray}
Equations \eq{8} and (9) are very difficult to solve.  However one
can discuss the hydrodynamic limit, where \eq{9a} is replaced by the
simpler constraint
\be
\label{10}
u^2(p,q,t) = u (p,q,t)
\ee
which says that $u(p,q,t)$ is a characteristic function consistent
with our idea of a classical fermi fluid.

Now consider a 2-dim. projection of the dynamics of the Liouville
equation \eq{8}, by introducing the moments:
\begin{eqnarray}
\label{11}
\rho (q,t) & = & \int^{+\infty}_{- \infty} {dp \over 2\pi}\: u(p,q,t)
\nonumber \\
\pi(q,t)\: \rho(q,t) & = & \int^{+\infty}_{- \infty} {dp \over 2\pi}\:
pu(p,q,t) \nonumber \\
\pi_2(q,t)\: \rho(q,t) & = & \int^{+\infty}_{-\infty} {dp \over
2\pi}\: p^2 u(\rho,q,t) \;\; {\rm etc.}
\end{eqnarray}
$\rho(q,t)$ and $\pi(q,t)$ correspond to the density and velocity of
the fluid.
The Liouville equation then corresponds to an infinite set of coupled
equations in 2-dim. involving the functions $\rho,\pi,\pi_2,\ldots$
which are constrained by \eq{10}\cite{eleven},
\be
\begin{array}{l}
\label{12}
\partial_t \rho + \partial_q (\rho \pi)  =  0. \\
\partial_t\pi = \partial_q \left(\displaystyle{{\pi^2 \over 2} + {q^2
\over 2}} - \pi_2\right) + \displaystyle{{\partial_q \rho \over \rho}}\:
\left(\pi^2 - \pi_2\right), \;
{\rm etc.}
\end{array}
\ee

If we further assume that the characteristic functions are
parameterized by \break
`quadratic' profiles which is a reasonable ansatz for
describing very small ripples on the fermi
surface:\cite{twelve,thirteen,fourteen}
\be
\label{13}
u(p,q,t) = \theta\left[(p_+(q,t) - p)\: (p - p_-(q,t))\right]
\ee
$p_\pm(q,t)$ parametrize the slightly deformed fermi surface in a way
that conserves fermion number.  Under the above approximation all the
moments of $u(p,q,t)$ depend only on the first two moments
$\rho(q,t)$ and $\pi(q,t)$, e.g. $\pi_2 = \pi^2 + {\pi^2 \over 3}\:
\rho^2$.  Substituting in
\eq{12} we get the closed set of field equations:
\begin{eqnarray}
\label{14}
\partial_t\rho + \partial_q (\pi\rho) & = & 0 \nonumber \\
\partial_t\pi + \pi\partial_q\pi & = & - \partial_q \left(- {q^2 \over 2}
+ {\scriptstyle {\pi^2 \over 2}}\: \rho^2\right)
\end{eqnarray}
The above are the hydrodynamic equations of a classical fermi fluid
with density $\rho(q,t)$, velocity $\pi(q,t)$ and pressure $P(q,t) = -
\displaystyle{ {q^2 \over 2}} +
{\scriptstyle {\pi^2 \over 2}}\: \rho^2(q,t)$.  They were originally obtained
using collective field theory\cite{twelve} rather than the
bosonization we have done.  Our method also brings out the fact that
collective field theory is an approximate but useful hydrodynamic
limit of the true bosonization of non-relativistic fermions.

The density and velocity satisfy the natural current commutation
relations: (which can be derived from the $W_\infty$ commutation
algebra of $u(p,q,t)$)
\be
\label{15}
\left[\rho(q,t),\pi(q',t)\right] = i\partial_q\delta(q-q')
\ee
Using \eq{15} one can see that the equation \eq{14} follow from the
action:
\begin{eqnarray}
\label{16}
S & = & \int dtdq \left(\pi\: {1 \over \partial_q}\: \dot\rho -
H(\rho,\pi)\right) \nonumber \\
H(\rho,\pi) & = & \int dq\: \left[{1 \over 2}\: \rho\left(\pi^2 + {1 \over
3}\: \rho^2\right) + \rho (V(q) + \mu)\right]
\end{eqnarray}
where $\mu$ is the lagrange multiplier corresponding to the constraint
$\int dq \rho = N$.  We can consider doing perturbation theory in the
action \eq{16} around the filled fermi sea where the density is $\rho_0
(q) = \sqrt{q^2 +2 \mu}, (\mu < 0)$:
\[
\rho(q,t) = \rho_0(q,t) + \partial_q \eta(q,t).
\]
Then introducing the `time of flight' variable $\tau$, by the relation
$\sqrt{\mu} \cosh \tau = q$ we get the action \eq{16} in the
chiral form.\cite{thirteen} (In ref. 13 this was obtained by computing
correlation functions using the underlying fermion theory.)
\begin{eqnarray}
\label{17}
S & = & \int d\tau dt \left(\partial_+ \eta \partial_- \eta
- {1 \over \mu}\: {1 \over \sinh^2 \tau}\: \left[(\partial_+ \eta)^3
- (\partial_- \eta)^3\right] + \ldots \right) \nonumber \\
\partial_\pm & = & \partial_\tau \pm \partial_t.
\end{eqnarray}
The range of $\tau$ in \eq{17} is $(0,-\infty)$, and the boundary
condition $\eta(0,t) = 0$ follows from \eq{9b}.  The action \eq{17} is
ill defined as the interaction diverges as $\tau \rightarrow 0$, which
is the turning point of the classical motion.  We have the problem of
``wall scattering''.  Using standard WKB methods we can replace
\eq{17} by an action in which the range of $\tau$ is $(-\infty +
i\epsilon, +\infty + i\epsilon), \epsilon > 0$.  In fact $\epsilon$
can be chosen to be $\pi \over 2$ and the $S$-matrix can be
calculated from the action
\be
\label{18}
S = \int^{+\infty}_{-\infty} d\tau dt\: \left(\partial_+\eta \partial_-
\eta - {1 \over \mu}\: {1 \over {\rm cost}^2\tau}\:
\left[\left(\partial_+\eta\right)^3 -
\left(\partial_-\eta\right)^3\right] + \ldots \right)
\ee
and then reinterpreted for ``wall scattering''.\cite{fifteen}  We
remark that \eq{18} has a natural interpretation in the momentum
representation on non-relativistic fermions, which naturally leads to
a non-singular interaction at $\tau = 0$.  For details we refer to
ref. 15.

\vskip 1cm
%\newpage

\noindent{\bf 3. The $S$-matrix}

\vskip 0.5cm

 From \eq{18} we can calculate the scattering amplitude of 4-massless
particles of energies $E_i$\cite{fourteen,fifteen,sixteen,seventeen}
\be
\begin{array}{l}
\label{19}
{\cal S} (1,2,3,4) \propto \delta(E_1 + E_2 + E_3 + E_4)\:
A(E_1,E_2,E_3,E_4) \\ [2mm]
A(E_1,E_2,E_3,E_4) = {1 \over \mu^2}\: (|E_1 + E_2| + |E_1 + E_2| +
|E_1+E_4| - i)
\end{array}
\ee
In the above we have used the mass shell conditions: $E_i + k_i = 0$.

Now we can interpret the above amplitude as a ``wall scattering''
process.  For example consider the ``$3 \rightarrow 1$'' process, in
which we scatter 3-particles at the ``wall'' and one comes back.  Let
us denote the \underbar{momenta} by $w_i$ with the identification
$w_1 \rightarrow -E_1, w_2 \rightarrow -E_2, w_3 \rightarrow -E_3$ and
$w_4 \rightarrow E_4$, then the momentum non-conserving ``wall
scattering'' amplitude is given by
\be
\label{20}
\bar A (1+2+3 \rightarrow 4) \propto {1\over \mu^2}\: \left(|w_1+w_2|
+ |w_1+w_3| + |w_1 - w_4| - i \right).
\ee
Connection with the ``wall scattering'' amplitude of 2-dim. string theory
can be made by the suitable multiplication of \eq{20} by leg pole
factors:\cite{eighteen,nineteen}
\[
\ell_{\rm in} (w) = \left({\pi \over 2}\right)^{i{w\over 4}} {\Gamma (iw)
\over \Gamma(-iw)}
\]
for incoming particles and
\[
\ell_{\rm out} (w) = \left({\pi \over 2}\right)^{-i{w\over 4}} {\Gamma
(-iw) \over \Gamma(iw)}
\]
for out-going particles.  The leg-pole factors have poles at imaginary
momenta: $w = -in$, reflecting the ``discrete states'' of the 2-dim.
string theory.\cite{twenty,twentyone}

Hence using \eq{20} the scattering amplitude for the process $1+2+3
\rightarrow 4$ is
\be
\label{21}
A(1+2+3 \rightarrow 4) = \left({\pi \over 2}\right)^{-{iw_4 \over 2}}
\prod^4_{i=1} {\Gamma(iw_i) \over \Gamma (-iw_i)}. \; \bar A(1+2+3
\rightarrow 4).
\ee
In a similar fashion one can obtain the amplitudes $A(1+2 \rightarrow
3+4)$, $A(1 \rightarrow 2+3+4)$ etc.

It is interesting to note that the scattering amplitudes can also be
derived directly from the $W_\infty$ conservation
laws.\cite{twentytwo,eighteen} Let us briefly mention these
conservation laws which easily follow from Liouville's equation \eq{8}:
\be
\label{22}
W_{rs} = \int {dpdq \over 2\pi} \: e^{-t(s-r)} (p-q)^r\: (p+q)^s\:
u(p,r,t).
\ee
One can check that ${d \over dt}\: W_{rs} = 0$.  The amplitude $\bar
A(1+2+3 \rightarrow 4)$ can be deduced by evaluating the charge
$W_{40}$ at times $t \rightarrow \pm \infty$, and then by obtaining a
relation between the incoming $(t \rightarrow - \infty)$ and outgoing
waves $(t \rightarrow +\infty)$.  The conserved charges \eq{22} are
the classical limit of the exactly conserved operators in which the
integrand $(p-q)^r (p+q)^s$ is replaced by phase space function
corresponding to the Weyl-ordered single particle operator $:(\hat p -
\hat q)^r (\hat p + \hat q)^s:$, where $::$ stands for Weyl ordering.
Hence the conserved operators get corrections in powers of
$g_{str}$.\cite{six} These quantum conserved charges can presumably be
used to compute corrections to the $S$-matrix in powers of the string
coupling.

\vskip 1cm

\noindent {\bf 4. Beyond Scattering Amplitudes: Classical Solutions}

\vskip 0.5cm

In the previous section we outlined a procedure to obtain all tree
level scattering amplitudes of the 2-dim. string theory in which the
external states are the massless tachyons.  However it is always more
fruitful to have a classical action from where we can derive the
scattering amplitudes.  But it is not very easy to write down such
an action as the scattering amplitudes only involve the massless
tachyon.  A different approach is to start with the continuum string
theory and use the $\sigma$-model approach.  In that approach it is
natural to introduce vertex operators that correspond to not only the
tachyon but also the graviton, dilaton etc.

The $\sigma$-model is described by the lagrangian
\be
\label{23}
{\cal S} = {1 \over 8\pi} \int d^2\xi\: \left({1 \over 2} \sqrt{\hat
g}\: \hat g^{ab} G_{\mu \nu}\partial_a X^\mu \partial_b X^\nu - 2
\hat R^{(2)} \Phi (x^\mu) + T(x^\mu) + \ldots \right)
\ee
$G_{\mu\nu},\; \Phi$ and $T$ correspond to the graviton, dilaton and
tachyon respectively.  We will consider a truncated $\sigma$-model and
presently ignore the other higher tensor fields.  $\hat g_{ab}$ is a
fiducial 2-dim. metric.  The standard equations of motion at one-loop
follow from the requirement of Weyl invariance: $\hat
g_{ab} \rightarrow e^\sigma \hat g_{ab}$\cite{twentythree}
\be
\begin{array}{l}
\label{24}
R_{\mu\nu} - 2\nabla_\mu \nabla_\nu \Phi + \nabla_\mu T \nabla_\nu T =
0 \\ [2mm]
R + 4(\nabla \Phi)^2 - 4\nabla^2 \Phi + (\nabla T)^2 + V(T) -4 = 0 .
\\ [2mm]
-2 \nabla^2 T + 4\nabla \Phi \cdot \nabla T + V'(T) = 0 \\ [2mm]
V(T) = - T^2 + \lambda T^3,
\end{array}
\ee
It is easy to see that these equations follow from the action
corresponding to 2-dim. dilaton-gravity coupled to $T$:
\be
\label{25}
S = \int d^2x\: e^{-2\Phi} \sqrt G \: \left(R - 4 (\nabla\Phi)^2 +
(\nabla T)^2 + V(T) - 4 \right)
\ee
In the above equations we have set $\alpha' = 2$.

\smallskip

\noindent \underbar{Classical Solutions}:

\smallskip

In the absence of the tachyon field $(T=0)$ one can exactly solve
\eq{24}.

The 1-parameter solution is given by
\begin{eqnarray}
\label{26}
ds^2 & = & G_{\mu\nu} dx^\mu dx^\nu = {dudv \over uv+a} \nonumber \\
e^{-2\Phi} & = & uv+a \nonumber \\
T & = & 0
\end{eqnarray}
where $u = t+x, v= t-x$. \\ [2mm] The above solution represents a
2-dim.  black-hole,\cite{twentyfour,twentyfive} if $a > 0$ and then `a'
can be identified with the mass of the black-hole.  One can compute
the scalar curvature corresponding to the metric in \eq{26},
\be
\label{27}
R = {4a \over uv+a}
\ee
The horizon is at $uv=0$ and the curvature singularity is on the
hyperbola $uv+a = 0$.  When $a=0$, \eq{26} corresponds to (after a
change of co-ordinates) to a flat space-time with a linear dilaton
backgrounds,
\begin{eqnarray}
\label{28}
G_{\mu\nu} & = & \eta_{\mu\nu}, \;\;\;\;  \Phi = {Q\eta \over 2}
\nonumber \\ [2mm]
Q & = & \sqrt 8
\end{eqnarray}

A few comments are in order:

\begin{enumerate}
\item[{(i)}] The black hole solution \eq{26} is remarkably similar to
the Schwarzschild solution of 4-dim. general relativity.  The hope is
that its occurrence, in a `soluble model' of string theory, may
eventually shed some light on some of the important unsolved problems
involving 4-dim.  black-holes.

\item[{(ii)}] The equation for tachyon propagation in the black hole
is easily derived from \eq{24} (neglecting non-linearities in the
tachyon potential):
\be
\label{29}
\left[4(uv+a)\: \partial_u\partial_v + 2 (u\partial_u + v\partial_v)
+ 1 \right]\; T(u,v) = 0.
\ee
This equation can be exactly solved but for our present purposes it is
only necessary to note that the solution has a logarithmic singularity
at the black-hole singularity:
\be
\label{30}
T(u,v) \sim \ln(uv + a).
\ee
\item[{(iii)}] Inclusion of higher order corrections in $\alpha'$ to the
$\beta$-function does not change the essential nature of the
black-hole solution.\cite{twentysix}

\item[{(iv)}] The solution \eq{28} is a starting point for a
recursive solution of the full set \eq{24}.  In fact \eq{28} produces
a tachyon background (neglecting non-linearities) $T_0(x) = (a+bx)\:
e^{{Q \over 2}\: x}$.  This in turn leads to a modification of the
background \eq{28} by terms of $o(e^{Qx})$ and so on.  It turns out
that tachyon scattering in the backgrounds $T_0(x) \sim o(e{Qx \over
2})$ and $G_{0\mu\nu} - \eta_{\mu\nu} \sim o (e^{Qx}),\: \phi_0 -
{Q \over 2}\: x \sim o (e^{Qx})$ reproduces the correct 3-point
amplitude, say $1+2 \rightarrow 3$.\cite{eighteen} The physical
picture is of wall scattering, the wall being provided by the above
mentioned backgrounds.  To compute a process like $1+2+3 \rightarrow
4$ (see eqn. \eq{21}) in this method one would presumably not only
have to solve for the backgrounds to higher orders in $e^{Qx}$ but may
also include the higher tensor fields of the string field theory.
This technology is beyond present capabilities and hence even though
the equations of motion provide the physical space-time picture, the
matrix model is the only computational tool we have.

\item[{(v)}] The $\beta$-function equations provide a natural framework
to study black-hole formation and evaporation.  However this has not
been possible because of the extreme difficulty of solving these
equations.  CGHS\cite{twentyseven} invented a simpler field theory
model by simply replacing $T$ in \eq{24} by a conformal scalar field.
In fact they introduced $N$ such scalar fields to develop a
semi-classical $1 \over N$ expansion for black-hole dynamics.  Though
we have learnt much along these lines it is fair to say that the basic
issues involving black-hole dynamics remain unresolved in that model.
\end{enumerate}

\vskip 1cm

\noindent {\bf 5. 2-dim. Black-Holes and the Matrix Model:}
\vskip 0.5cm

In the remaining part of this note we will focus on the possibility of
discussing properties of black-holes in the 2-dim. target space using
the matrix model.\cite{twentyeight}  For other approaches we refer to
refs. (29) and (30).

Any attempt to understand the emergence of a non-trivial space-time in
the matrix model has to contend with the fact that non-relativistic
fermions are formulated in a flat space-time.  However as we have seen
the ``space-time'' in which the small perturbations on the fermi
surface propagate is the half-plane with perfectly reflecting boundary
conditions and local interactions.  On the other hand since the target
space theory has a metric $G_{\mu\nu}$ and a dilaton $\Phi$, one can
imagine an equivalent description of the system in which there is a
field redefinition of the metric: $\tilde G_{\mu\nu} = G_{\mu\nu}
e^{-2\Phi}$, which corresponds to a space time which is flat, at least
for the solution \eq{27}.  Of course the field redefinition of the
metric will imply in general a non-local and non-linear redefinition
of the tachyon field.

In the following we present a transformation of the quantum phase
space distribution of fermions, $\hat u(p,q,t)$ which was defined in
eq. \eq{7}.
\be
\label{31}
\hat\phi (p,q,t) = \int dp'dq'\: K(p,q|p',q')\; \hat u(p',q',t)
\ee
where
\be
\label{32}
K(p,q|p',q') = |(p-p')^2 - (q-q')^2|^{-1/2}
\ee
The equation of motion for $\hat u$, that follows from the
fermion equation of motion, is $\partial_t\hat u + p\partial_q
\hat u + q\partial_p\hat u = 0$.  One can verify that
$\hat\phi$ satisfies the same equation and hence if we introduce the
variables $u = {1 \over 2}\: e^{-t} (p+q),\; v = {1\over 2}\:
e^t(p-q)$, we see that the equation of motion imply
$\partial_t\hat u (ue^t + v e^{-t}, ue^t - v e^{-t}, t) = 0$
and $\partial_t \hat \phi (ue^t + v e^{-t}, ue^t - v e^{-t}, t) =
0$.  Hence \eq{31} effectively becomes a 2-dim. relation.  Defining
$\hat T(u,v) = \hat \phi (u+v,u-v,0)$ we have
\begin{eqnarray}
\label{33}
\hat T(u,v) & = & \int du'dv' \tilde K(u,v|u',v')\: \hat u
(u'+v', u'-v',0) \nonumber \\
\tilde K &=& |(u-u')\: (v-v')|^{-1/2}
\end{eqnarray}
Now if \eq{33} has anything to do with black-holes it should have the
property that for low energy scattering the background metric and
dilaton perceived by the field $\hat T(u,v)$, correspond to the
classical solution \eq{27}.

We demonstrate this in two steps:  First, consider a state $|\psi>$ in
the fermion theory which differs from the classical ground state
$|\psi_0>$ so that $\delta u(p,q,t) = \langle \psi|\hat u
(p,q,t)|\psi\rangle - \langle\psi_0|\: \hat u(p,q,t)|\psi_0\rangle$
has support, at most in a small neighbourhood of the fermi surface
$p^2-q^2-2\mu = 0$.  Then $\delta T(u,v) = \langle \psi|\hat
T(u,v)|\psi\rangle - \langle \psi_0|\hat T(u,v)|\psi_0\rangle$, is given by
\be
\label{34}
\delta T(u,v) = \int du'dv' \hat K(u,v|u',v')\: \delta u(u',v')
\ee
The second step is that $\tilde K(u,v|u',v')$ has the following
property:
\be
\label{35}
\left[4\left(uv - {\mu \over 2}\right)\: \partial_u\partial_v + 2
(u\partial_u + v\partial_v) + 1 \right]\; \tilde K(u,v|u',v') = o
\left(u'v' - {\mu \over 2}\right).
\ee
Now since $\delta u(u',v')$ has support in a small region around the
fermi surface $u'v' = {\mu \over 2}$, \eq{34} and \eq{35} imply that
\be
\label{36}
\left[4\left(uv - {\mu \over 2}\right)\: \partial_u\partial_v + 2
(u\partial_u + v\partial_v) + 1 \right]\; \delta T(u,v) =  o
\:\left({\delta E \over \mu}\right) \approx o
\ee
The differential operator on the l.h.s. in \eq{36} is precisely the one
that occurs in \eq{29} with the identification $a = - {\mu \over 2}$.
Hence the fermi level, the only dimensional parameter that specifies
the ground state, is identified with the black-hole mass.  $\delta E$
in \eq{36} is the maximum energy of the fermi fluid in the region
deformed from the filled fermi sea.  $\delta E/\mu$ is the expansion
parameter proportional to the string coupling.

It is important to emphasize that even though \eq{34} is a non-local
linear functional of $\delta u(u,v)$, it is a non-local and non-linear
function of the collective fields $p_\pm (q,t)$ that were introduced in
\eq{13}.  In fact one can explicitly express it as a power series in
the fluctuations $\eta_\pm(q,t) = p_\pm (q,t) - p^0_\pm (q), p^0_\pm
(q) = \pm \sqrt{q^2 + 2\mu}$.  This fact makes it a difficult
enterprise to derive a closed equation for $\delta T(u,v)$.

\vskip 1cm

\noindent {\bf 6. The Question of Singularities in the
Black-Hole Background}

\vskip 0.5cm

In the previous section we had stated that the tachyon field
propagating in the black-hole background develops a singularity (eqns.
\eq{29},\eq{30}), where the space-time curvature is singular.  This
would be one way of perceiving the black-hole singularity in an
effective theory of tachyons.  We now demonstrate that $\delta T(u,v)$
as defined in \eq{34} using the fermi fluid theory has no singularity
at $uv - {\mu \over 2} = 0$.

The issue is that of a 2-dim. integral of the form $\int dx dy
f(x,y)|x^2-y^2|^{-{1\over 2}}$.  It is clear that such an expression
is singular only if the function $f(x,y)$ is singular, because
$|x^2-y^2|^{-{1\over 2}}$ $dxdy$ is regular at $x=y=0$.  Now in our
theory $\delta u(u',v')$ is non-singular simply because it is a
difference of 2 characteristic functions: $\delta u = u-u_0$.  More
generally, in the full quantum theory, a general state of the fermion
theory $|\psi>$ is obtained as a $W_\infty$ rotation of the fermion
ground state $|\psi_0>$.  This implies that $\langle \psi|\hat
u(p,q,t)|\psi \rangle$ is obtained as a $W_\infty$ rotation of the
regular function $\langle \psi_0|\hat u(p,q,t)|\psi_0
\rangle$, and hence $\langle\psi|\hat u(p,q,t)|\psi \rangle$ is a
regular function on phase space.

The above general agreement can be supplemented by an explicit
calculation in the case when $\delta u(u,v)$ is a simple local
deformation of the fermi-surface.  We simply quote the formulae from
the literature.\cite{twentyeight} Consider $\delta u(u,v)$ so that it
is non-zero in a `small' region around the fermi surface $uv = {\mu
\over 2}$.  Here by small we mean that the maximum energy of fermions
in that region is given by $E = \mu +
\Delta$, and $\Delta
\ll |\mu|$.  $\Delta \over \mu$ is proportional to the string
coupling.  Then a calculation gives
\begin{eqnarray}
\label{37}
\delta T(u,v) & \simeq& \left(- {\mu \over 2}\right)^{-{1\over 2}}
\left[\left(uv - {\mu \over 2}\right)\: \ln |uv - {\mu \over 2}| -
\left(uv - {\mu + \Delta \over 2}\right)\: \ln |uv - {\mu + \Delta \over
2}|\right] \nonumber \\ [2mm]
&+& {\rm (regular~terms)}
\end{eqnarray}
It is clear that $\delta T(u,v)$ has no
singularities.  It is regular at $uv = {\mu \over 2}$ and $uv = {\mu +
\Delta \over 2}$.  {\em However an expansion of \eq{37} in powers of
${\Delta \over |\mu|}$ is divergent at $uv = {\mu \over 2}$ in every
order of perturbation theory}.
\be
\label{38}
\Delta T(u,v) \simeq |{\Delta \over \mu}| \ln |uv - {\mu \over 2}| +
|{\Delta \over \mu}|^2 \left(uv - {\mu \over 2}\right)^{-1} + o\left(|{\Delta
\over \mu}|^3\right)
\ee
The first term is the leading logarithmic divergence which we have
already encountered in \eq{30}.  It is clear that the non-linear
completion of eqn. \eq{29}, that arises from string theory, cures the
singularity if one sums \eq{38} to all orders in perturbation theory.
In that sense \eq{37} is actually a non-perturbative result.

\vskip 1cm

\noindent{\bf 7.\footnote{This section has been added for
completeness in the proceedings and was not part of the talk.}
Boundary condition and Tachyon Propagation in black-hole
background}\cite{thirtyone}

\vskip 0.5cm

The matrix model is exactly soluble; however the main difficulty in the
subject is the space-time interpretation of the answers that emerge
from the matrix model.  Presently the only known way is to invent
relevant transformations between quantities in the matrix model and
2-dim.  string theory.  Since the latter is mainly formulated in
perturbation theory, we can compare the two theories at the classical
level.  This was the spirit behind the definitions \eq{21} and
\eq{34}.

The results that we presented about tachyon propagation, in a
black-hole background in 2-dim.  string theory via the matrix model
made very minimal assumptions on the function $\delta u(u,v)$.  We now
discuss the boundary conditions on $\delta T(u,v)$, defined by
\eq{34}, that correspond to various processes involving the scattering
of tachyons by a black-hole.  Firstly we note that $\delta T(u,v)$
defined by \eq{34} is non-zero in the entire Kruskal plane $P =
\{(u,v)| - \infty < (u,v) < +\infty \}$.  In particular
a generic $\delta u(u,v)$ gives, besides incoming flux from ${\cal I}^-$,
a flux emerging from the white hole (see Fig.1).  The demonstration
of non-singular propagation would be more relevant for a more
realistic collapse scenario if we could by some means avoid a flux of
particles emanating from the line $v = 0$ in the Kruskal plane, and
consider the half plane $P_+ = \{(u,v)| v \geq 0\}$ as the physical
space-time along with the boundary condition $\delta T(u,v = 0) = 0$.

A sufficient condition that achieves this boundary condition is
\be
\label{39}
\int^\infty_{-\infty} dv'|v'|^{-{1\over 2}} \delta u(u',v') = 0
\ee
We will show that for a very large class of fluctuations
$\delta u_1(u,v)$ it is possible to construct a modified fluctuation
$\delta u(u,v) = \delta u_1(u,v) - \delta u_2(u,v)$ (where $\delta
u_2(u,v)$ depends on $\delta u_1(u,v)$) which satisfies \eq{39}.  We
can use this result to generate infinitely many solutions of \eq{39}.

We will show the result (easily generalizable to other cases) for
$\delta u_1(u,v)$ which consists of one ``blip'' bulge in the fermi
surface, and one ``antiblip'' (dip in the fermi surface).  Further we
will assume that the modified fermi surface has a ``quadratic
profile''.  This means, e.g., that the blip is described by the
formula
\be
\label{40}
\delta u_1 (u,v) = \theta\left[(v_+(u) - u)\: (u - v_-(u))\right].
\ee
between some $u_1$ and $u_{\rm max}$ (see Fig. 2).  In eq. \eq{40} and
Fig. 2 we have ignored the antiblip which can be discussed similarly.
The figure shows that $v_+(u) = v_+^{(0)} (u) = \displaystyle {\mu \over
2u}$ for $u\:\epsilon\: [u_1,u_2]$.  Let us consider a fluctuation
\[
\delta u_2(u,v) = \theta \left[(\tilde v_+(u) - v)\: (v - \tilde
v_-(u))\right]
\]
for $u\:\epsilon\: [u,u_{\rm max}]$ and zero elsewhere.  Clearly \eq{39}
can be satisfied by $\delta u(u,v)$ if
\begin{eqnarray*}
\sqrt{\tilde v_+(u)} & = & \sqrt{v_+(u)}\: - f(u) \\
\sqrt{\tilde v_-(u)} & = & \sqrt{v_-(u)} + f(u)
\end{eqnarray*}
where $f(u)$ is arbitrary and positive in the region $[u_2,u_{\rm
max}]$ and vanishes elsewhere.

Note that $\delta u_1(u,v)$ and $\delta u_2(u,v)$ can, like before, be
chosen close enough to the fermi surface so that eq. \eq{36} is again
valid. Note that $\delta u(u,v)$ that we have constructed is not a
``quadratic profile'' and in the range $(u_2,u_{\rm max})$, a $u = {\rm
constant}$ line intersects it in 4-points.

The point of the above demonstration is that if we use fermi fluid
distributions that eliminate the particle flux from the white hole,
then our previous demonstration of the absence of singularity in
$\delta T(u,v)$ is more relevant to the more physical collapse
scenarios of black-hole dynamics.  Solutions like the one described
above correspond to nomalizable wave packets at ${\cal I}^-$ and
${\cal I}^+$.

We also want to state that collapse scenarios for black
holes are bound to be different in case the underlying theory (like in
the case of 2-dim. string theory) has an infinite number of conserved
charges.

\vskip 0.5cm
\noindent{\bf Concluding Remarks}:
\vskip 0.5cm

We have presented a certain view of 2-dim. string theory.  If one
takes stock of the achievements it is fair to say that the matrix
model and the leg pole prescription enables in principle a calculation
of the $S$-matrix for wall scattering to all orders in perturbation
theory.  An explicit demonstration of higher order corrections to the
$S$-matrix using the quantum (conserved) $W_\infty$ charges would be
desirable

Regarding non-perturbative effects, we could demonstrate that the
absence of a singularity in the tachyon wave at the black hole
singularity was indeed a non-perturbative effect.  This is because one
needs to sum a series each term of which is divergent at every order
of the semi-classical expansion, at the black hole singularity, but
the full sum is finite and singularity free.

Regarding open questions we would list the following:

\begin{enumerate}
\item[{(i)}] We do not yet know what the stringy non-perturbative
effects\cite{thirtytwo} $\left(\sim e^{-{1 \over g_{str}}}\right)$
means in the target space of 2-dim. string theory.  In a manner of
speaking we know the answer but not the question.

\item[{(ii)}] The same comment also applies to the strongly coupled
2-dim. stringy theory.  In the matrix model this means $\mu = {1 \over
g_{str}} \rightarrow 0$, and the fermi level is very near the tip of
the inverted harmonic oscillator potential.  Clearly the picture of
``wall scattering'' is not applicable anymore as the fermi fluid can
easily trickle to the `other side' of the potential.  We do not know
the `strong coupling' question in 2-dim. string theory.  The situation
here is more difficult than it was in gauge theories in the 1970s,
because at that time lattice gauge theorists had a phenomenological
picture of quark confinement that they wanted to explain: The
squeezing of chromo-electric flux between a quark and anti-quark.  It
would be interesting to know the corresponding questions in string
theory in two or for that matter even in four dimensions.

\item[{(iii)}] It is conceivable that the $\beta$-function equations
\eq{24} and their quantization can describe the formation and
evaporation of black holes in 2-dim. string theory.  It would be of
great interest to know how these processes can be described in the
matrix model.

\end{enumerate}

\vskip 0.5cm
\noindent {\bf Acknowledgement}:
\vskip 0.5cm

The `view' of 2-dim. string theory and black hole physics presented
here was mainly developed with Avinash Dhar and Gautam Mandal and also
in the early stages with Sumit Das and Anirvan Sengupta.  I have
greatly enjoyed the experience and and I thank them for it.
I would like to thank Gautam Mandal for a careful reading of the
manuscript.

\vskip 1cm
%\noindent{\bf References}
%\vskip 0.5cm

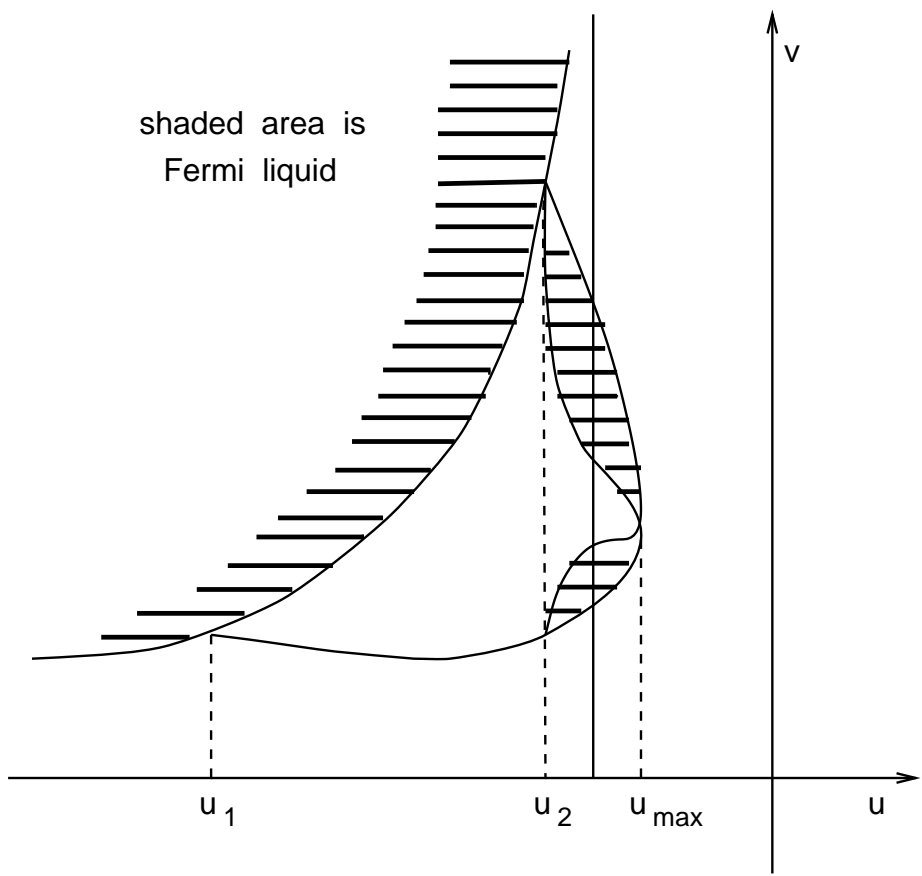
\end{document}